# Novel neural-network architecture for continuous gravitational waves

Prasanna M. Joshi[1,2,*] and Reinhard Prix[1,2]

[1]*Max Planck Institute for Gravitational Physics (Albert-Einstein-Institute), 30167 Hannover, Germany*
[2]*Leibniz Universität Hannover, 30167 Hannover, Germany*

The high computational cost of wide-parameter-space searches for continuous gravitational waves (CWs) significantly limits the achievable sensitivity. This challenge has motivated the exploration of alternative search methods, such as deep neural networks (DNNs). Previous attempts [1,2] to apply convolutional image-classification DNN architectures to all-sky and directed CW searches showed promise for short, one-day search durations, but proved ineffective for longer durations of around ten days. In this paper, we offer a hypothesis for this limitation and propose new design principles to overcome it. As a proof of concept, we show that our novel convolutional DNN architecture attains matched-filtering sensitivity for a targeted search (i.e., single sky-position and frequency) in Gaussian data from two detectors spanning ten days. We illustrate this performance for two different sky positions and five frequencies in the 20–1000 Hz range, spanning the spectrum from an "easy" to the "hardest" case. The corresponding sensitivity depths fall in the range of $82\text{--}86/\sqrt{\text{Hz}}$. The same DNN architecture is trained for each case, taking between 4–32 hours to reach matched-filtering sensitivity. The detection probability of the trained DNNs as a function of signal amplitude varies consistently with that of matched filtering. Furthermore, the DNN statistic distributions can be approximately mapped to those of the $\mathcal{F}$-statistic under a simple monotonic function.

## I. INTRODUCTION

Continuous gravitational waves (CWs) are weak, long-lasting and nearly-monochromatic waves emitted by non-axisymmetric spinning neutron stars. Numerous searches have been performed on the data from the LIGO (H1 and L1) and Virgo (V1) detectors, yet no CWs have so far been detected [3]. Owing to the expected small amplitude of CW signals, months to years of data will be required in order to collect a sufficient signal-to-noise ratio to allow a detection.

The most sensitive search method consist of *coherently* integrating signal templates over the entire time-span of the data, as process commonly known as *matched filtering*. However, for wide parameter spaces this method is severely constrained by the required computing cost (due to the astronomical number of required templates), e.g., see [4]. Instead, *semicoherent* search methods are used in practice for wide parameter-space searches, which proceed by coherently analyzing shorter segments of data and combining their results incoherently. These search methods tend to result in the highest sensitivity at a fixed computational cost.

In order to overcome the computational-cost constraint on the achievable sensitivity, one alternative approach being explored is to use deep neural networks (DNNs) to search for CWs in the data. There has been a number of studies exploring the potential of DNNs to help improve CW searches, for example, as a clustering and follow-up method of search candidates [5–7], to reduce the computational cost of follow-ups [8,9], and to mitigate the effect of instrumental noise artifacts [10]. DNNs have also been shown to be able to accelerate searches for long-duration, transient CWs [11,12].

Here we continue to pursue the approach of [1,2] to train DNNs as an *end-to-end* search method for CW signals directly on the detector strain data. These earlier studies considered wide-parameter-space searches of signals with time-spans of $10^5$ s and $10^6$ s, respectively, using the best available DNN architectures for image classification tasks. While this approach proved quite effective on the shorter time-span of $10^5 \text{ s} \sim 1$ day, it performed poorly on the longer signals of $10^6 \text{ s} \sim 11.6$ days.

This raises the question if it is the larger number of signal waveforms (i.e., *templates*) in a wide parameter space, or the morphology of longer signals itself that thwarts the networks' ability to successfully learn to detect them. A similar difficulty of DNNs to detect long-duration signals has also been observed in the context of compact-binary-coalescence searches, see [13]. We find that the characteristics of longer CW signals seem to be the underlying cause of the inability of the image-classification architectures to effectively learn to detect them.

[*]Corresponding author: prasanna.mohan.joshi@aei.mpg.de

Therefore here we take a step back and focus on the problem of targeted (i.e., single-template) CW searches over a time-span of ten days in simulated Gaussian noise. Developing architectures capable of detecting longer CW signals is crucial for scaling up to complete wide-parameter space search methods that could ultimately compete with current state-of-the-art semicoherent CW searches.

The plan of the paper is as follows: in Sec. II, we introduce the CW signal model, we define benchmark targeted search cases in Sec. III and describe the architecture and training of our DNN in Sec. IV. Finally, we present our test results and discussion in Sec. V and conclusions and future outlook in Sec. VI.

## II. CONTINUOUS GRAVITATIONAL WAVES

Continuous gravitational waves are long-lasting, quasi-monochromatic waves with a slowly varying frequency, emitted by spinning nonaxisymmetric neutrons stars. We model the evolution of the CW signal phase $\Phi(\tau)$ as a function of time $\tau$ in the source frame (assuming only linear spindown) as

$$\Phi(\tau) = 2\pi \left[ f(\tau_{\rm ref})\Delta\tau + \frac{1}{2}\dot{f}(\tau_{\rm ref})\Delta\tau^2 \right] + \phi_0, \quad (1)$$

where $\Delta\tau \equiv \tau - \tau_{\rm ref}$, and $\tau_{\rm ref}$ is the reference time at which $f(\tau_{\rm ref})$ and $\dot{f}(\tau_{\rm ref})$ are defined.

In the detector frame the signal experiences frequency modulation due to the relative motion between the detector and the source. This modulation can be characterized by the relation between the arrival time $t$ of a wave front at the detector that left the source at time $\tau$. For an isolated neutron star, this timing relation $\tau(t)$ can be written as follows:

$$\tau(t; \mathbf{n}) = t + \frac{\mathbf{r}(t) \cdot \mathbf{n}}{c} - \frac{d}{c}, \quad (2)$$

where $\mathbf{n} = (\cos\delta\cos\alpha, \cos\delta\sin\alpha, \sin\delta)$ is the unit vector pointing to the source in equatorial coordinates, expressed in terms of right ascension ($\alpha$) and declination ($\delta$), $\mathbf{r}(t)$ is the vector from the solar-system barycenter (SSB) to the detector location, $d$ is the distance between the SSB and the source and $c$ is the speed of light. The term $\mathbf{r}\cdot\mathbf{n}/c$ is known as the Rømer delay.

The frequency evolution $f(t)$ of the signal in the detector frame is obtained by applying the timing relation $\tau(t)$ of Eq. (2) to the source-frame phase evolution of Eq. (1), namely $\Phi(t) = \Phi(\tau(t))$, and computing the derivative

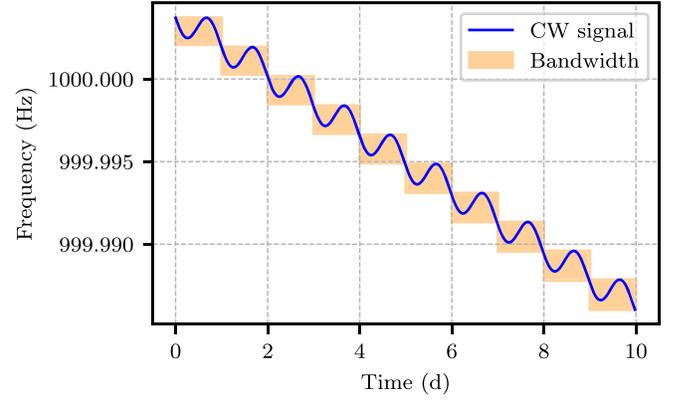

FIG. 1. Detector-frame frequency evolution $f(t)$ of Eq. (3) for a CW signal with source-frame frequency $f(\tau_{\rm ref}) = 1000$ Hz, spindown $\dot{f}(\tau_{\rm ref}) = -10^{-10}$ Hz s$^{-1}$ and sky position Sky-B (see Table I). The highlighted region denotes the bandwidth of the signal over each one-day time span.

$$f(t; \lambda) = \frac{d\Phi(t)}{2\pi dt} = [f(\tau_{\rm ref}) + \dot{f}(\tau_{\rm ref})\Delta\tau(t)]\frac{d\tau}{dt}, \quad (3)$$

where $\lambda \equiv \{f(\tau_{\rm ref}), \dot{f}(\tau_{\rm ref}), \alpha, \delta\}$ are commonly referred to as the phase-evolution parameters.

An example of the detector-frame frequency evolution $f(t)$ for a CW signal over a time-span of ten days is shown in Fig. 1. Here we see the two-component Doppler modulation of the signal due to the diurnal rotation of the detector (Doppler shifts of order $\sim 10^{-6} f$) and the orbital motion of the Earth (Doppler shifts of order $\sim 10^{-4} f$ over the course of a year).

The CW strain signal in the detector additionally depends on four *amplitude parameters* $\mathcal{A}$, namely the overall signal amplitude $h_0$, the neutron-star spin-axis alignment $\cos\iota$ with the line of sight, the polarization angle $\psi$ and the initial phase $\phi_0$. The full expression for the strain signal $h(t; \mathcal{A}, \lambda)$ is not important here and can be found, for example, in [4,14]. The total measured strain $x(t)$ in a detector can be expressed as

$$x(t) = n(t) + h(t; \mathcal{A}, \lambda), \quad (4)$$

where $n(t)$ denotes the noise, characterized by a noise power spectral density $S_n(f)$ as a function of frequency. In practice $n(t)$ is often assumed to be (approximately) Gaussian, a simplifying assumption that we will also use in this work.

We can distinguish three main categories of CW searches [3,4], depending on the assumed level of knowledge about the signals; wide parameter-space *all-sky* searches assume the phase-evolution parameters $\lambda$ to be completely unknown, *directed* searches treat the sky position $\mathbf{n}$ as known with unknown frequency and spindown(s), while *targeted* searches take the phase-evolution parameters $\lambda$ to be fully known. Note that the four amplitude parameters $\mathcal{A}$ are typically considered unknown even for targeted searches.

The *sensitivity* of a CW search [15,16] is typically characterized in terms of an *upper-limit* amplitude $h_0^{p_{\text{det}}}$ at which a search achieves a given detection probability $p_{\text{det}}$ (typically chosen as 90% or 95%) at a chosen false-alarm level $p_{\text{fa}}$. This upper-limit amplitude $h_0$ characterizes a *population* of signals with unknown (neutron-star) spin axis orientation (uniform priors $\cos\iota \in [-1, 1]$ and $\psi \in [-\pi/4, \pi/4]$) and initial phase (uniform in $\phi_0 \in [0, 2\pi]$). The CW upper limit amplitude $h_0^{p_{\text{det}}}$ scales with the amplitude noise spectral density $\sqrt{S_n}$ at every frequency, it is therefore more convenient to use the *sensitivity depth* $\mathcal{D}^{p_{\text{det}}}$, defined as

$$\mathcal{D}^{p_{\text{det}}} \equiv \frac{\sqrt{S_n}}{h_0^{p_{\text{det}}}}, \quad (5)$$

which characterizes the sensitivity of a search setup [16] independently of the noise-floor level $S_n$. In the following we use the sensitivity depth $\mathcal{D}^{90\%}$, corresponding to an upper-limit amplitude $h_0^{90\%}$, for which a matched-filter search would achieve a detection probability of $p_{\text{det}} = 90\%$ at a false-alarm probability of $p_{\text{fa}} = 1\%$.

## III. BENCHMARK TARGETED SEARCHES

Previous studies [1,2] had directly attempted to tackle wide-parameter-space CW searches with convolutional deep-neural-network architectures from image classification. While this approach worked well for short search durations of about one day, it became ineffective when extended to longer durations up to $10^6$ s $\sim 11.6$ days, see Table VI in [2].

Further experimentation reveals that these network architectures struggle with longer-duration signals even when trained on much simpler targeted searches. For example, for ten-day signals the Inception-Resnet-v2 architecture used in [2] fails to essentially learn anything except at the lowest frequency ($f = 20$ Hz), as shown in Fig. 2.

This indicates that it is not (only) the larger parameter space but the signal morphology itself that causes problems. Specifically, the difficulty encountered by DNNs in detecting CW signals appears to be correlated with their effective bandwidth in the data, corresponding to the Doppler broadening in the detector frame (as illustrated in Fig. 1).

In this study, we therefore narrow our focus on targeted searches spanning ten days, in order to demonstrate, as a proof of concept, that an appropriately-designed DNN architecture can detect these longer signals with optimal matched-filter sensitivity. For this purpose we define ten benchmark cases of targeted ten-day searches, given in Table I, spanning the spectrum from "easy" to "hardest", with five different frequencies from 20–1000 Hz (higher frequency leads to more Doppler broadening) and two different sky positions, Sky-A and Sky-B.

Sky position Sky-B has the widest Doppler broadening (over the sky) of the signal during the ten-day search span, while sky position Sky-A is more favorable with a narrow signal bandwidth. The total ten-day signal bandwidths for all benchmark cases are listed in Table II. We see that the search at $f = 20$ Hz targeting sky position Sky-A has the narrowest signal band (∼0.09 mHz), and is expected to be the easiest to master for a DNN, while targeting Sky-B at $f = 1000$ Hz is expected to be the hardest case (with a total bandwidth of ∼18.7 mHz). A visual illustration of the respective bandwidths of signals at the two sky-positions can also be found in Fig. 3.

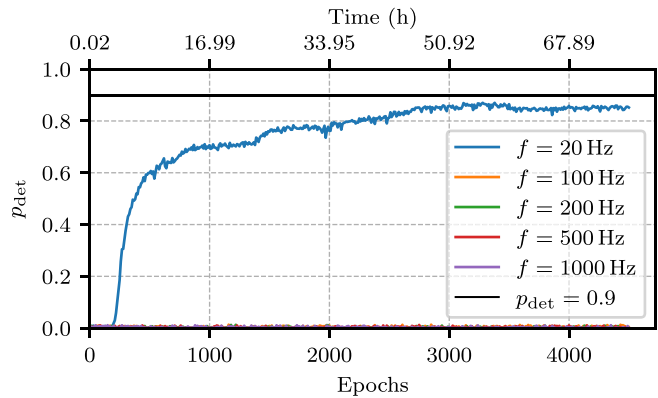

FIG. 2. Training progress of an image-classification DNN (Inception-Resnet-v2) used in [2] when trained on a ten-day targeted search considered in the current study (at sky position Sky-B, see Table I). Trained signals are injected at depth $\mathcal{D}^{90\%}_{\text{Sky-B}}$ given in Eq. (6) at five different frequencies. In 76 h of training, the image-classification network manages to learn to detect such signals only in the lowest-frequency case ($f = 20$ Hz), while completely failing for higher frequencies ($f \geq 100$ Hz).

TABLE I. Benchmark targeted searches, spanning five frequencies between 20–1000 Hz and two sky positions, one "easy" (Sky-A) and the "hardest" (Sky-B).

| | |
|---|---|
| Start time | 1200300463 s |
| Duration | 10 days |
| Detectors | LIGO Hanford (H1) and Livingston (L1) |
| Noise | Stationary, white, Gaussian |
| Frequency $f(\tau_{\text{ref}})$ | 20, 100, 200, 500, 1000 Hz |
| Spindown $\dot{f}(\tau_{\text{ref}})$ | $-10^{-10}$ Hz s$^{-1}$ |
| $\tau_{\text{ref}}$ | 1200300463 s |
| sky position $(\alpha, \delta)$ | Sky-A (6.123771, 1.026457) rad |
| | Sky-B (2.119314, 0.299076) rad |

TABLE II. Total signal bandwidths over ten days (in mHz) for the targeted-search cases defined in Table I.

| | Bandwidth [mHz] | |
|---|---|---|
| Frequency [Hz] | Sky-A | Sky-B |
| 20 | 0.089 | 0.453 |
| 100 | 0.163 | 1.941 |
| 200 | 0.382 | 3.802 |
| 500 | 1.060 | 9.383 |
| 1000 | 2.194 | 18.685 |

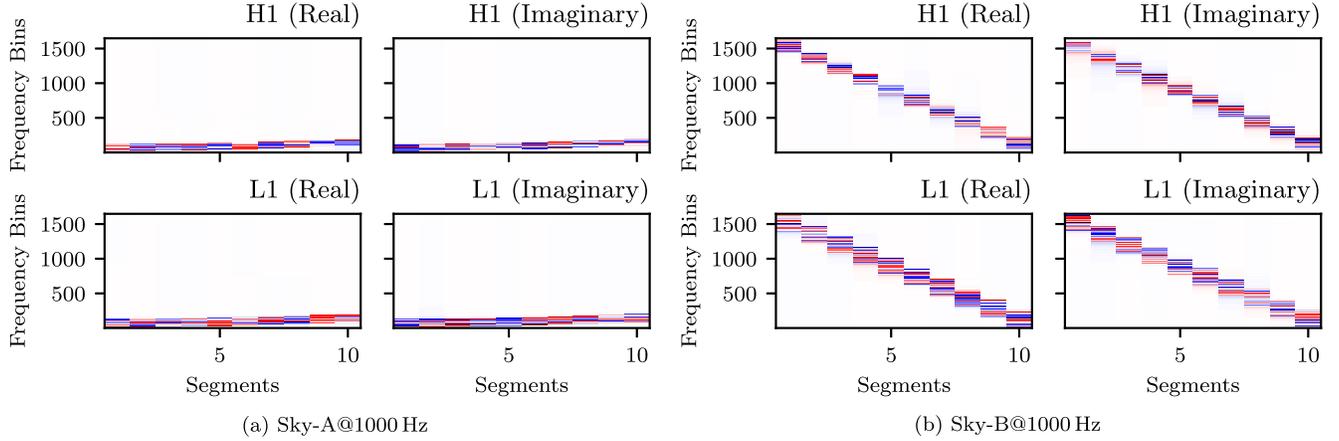

FIG. 3. Input spectrogram arrays for two signals without noise at 1000 Hz: (a) sky position Sky-A featuring low Doppler broadening of the signal, and (b) sky position Sky-B exhibiting maximal Doppler broadening over the ten-day period. See Table I for the complete parameter definitions.

We can estimate the optimal matched-filtering sensitivity $\mathcal{D}^{90\%}$ (at $p_{\text{fa}} = 1\%$ false-alarm level) for each of the benchmark search cases using the approach developed in [15,16] and implemented in [17].[1] The corresponding optimal sensitivity depths depend on the sky position (due to the different antenna-pattern response), and are obtained as

$$\mathcal{D}^{90\%}_{\text{Sky-A}} \approx 86.2 \ /\sqrt{\text{Hz}},$$
$$\mathcal{D}^{90\%}_{\text{Sky-B}} \approx 81.8 \ /\sqrt{\text{Hz}}. \quad (6)$$

This defines the optimal sensitivity ceiling to compare the DNN performance against.

Note that the matched-filter search at Sky-B is slightly less sensitive than at Sky-A, requiring a stronger signal (i.e., smaller depth) to reach $p_{\text{det}} = 90\%$. This is due to differences in the antenna-pattern response at the two sky positions and is unrelated to the previous discussion about signal bandwidths in the detector frame, which does not affect matched-filter performance.

## IV. DEEP LEARNING

In this section we describe the design of the DNN architecture, the pre-processing of the input data, and the training process.

---
[1]This is assuming the $\mathcal{F}$-statistic, which is not strictly Neyman-Pearson optimal [18], but the difference is too small to be of practical relevance for these search setups.

### A. A new DNN architecture for CWs

Deep state-of-the-art image classification networks (specifically, ResNet [19] and Inception-ResNet-v2 [20]) employed in [1,2] were unable to achieve competitive sensitivities for CW signals lasting ∼11.6 days, with rapidly decreasing performance at higher frequencies (e.g., see Table VI in [2]). As mentioned in the previous section, these architectures perform poorly even when simplifying the problem to simple targeted searches over ten days, as seen in Fig. 2.

We hypothesize that this failure to learn is due to a mismatch between the *morphology of long CW signals* in noise and the *implicit priors* in (convolutional) image-classification network architectures. These image-classification priors can be roughly characterized as

(i) the image could represent any object,
(ii) high signal-to-noise-ratio pixels can be combined locally to find small-scale structures like ridges, corners, etc.,
(iii) lower-level patterns can be hierarchically combined into larger structures in subsequent layers, where the exact location of lower-level structures has little to no impact on the final classification.

The resulting typical convolutional image-classification architectures consists of small convolutional kernels (such as $7 \times 7$ or smaller), lossy layer reductions such as max- or mean-pooling and a large number ($\gtrsim 50$) of layers (cf. [19,20]).

Contrast this with CW signals, where two things happen when increasing search duration: (i) the signal depth $\mathcal{D}^{90\%}$ for matched filtering grows as $\propto \sqrt{\text{duration}}$, so the amplitude of the target signals become smaller, and (ii) the Doppler spreading of the signal in the detector frame increases, see Sec. III and Fig. 1. Both of these factors contribute to *weaker* and *less localized* signal power in time-frequency space, i.e., reduced local signal-to-noise ratio in any spectrogram bin. Figure 4 illustrates this effect for a spectrogram of the Sky-B@1000 Hz signal of Fig. 3 added to Gaussian noise at a matched-filtering depth of $\mathcal{D}^{90\%} = 81.8/\sqrt{\text{Hz}}$. The resulting "image" does not show any visible trace of the signal. We can therefore roughly characterize the *CW signal priors* as:

(i) a diurnal narrow frequency pattern, repeating daily with an overall frequency drift (due to spindowns and orbital motion[2]), see Figs. 1 and 3,
(ii) a vanishing local signal-to-noise ratio in any spectrogram pixel, see Fig. 4,
(iii) a lossless combination of *all* signal pixels will be necessary for classification to be able to compete with matched filtering.

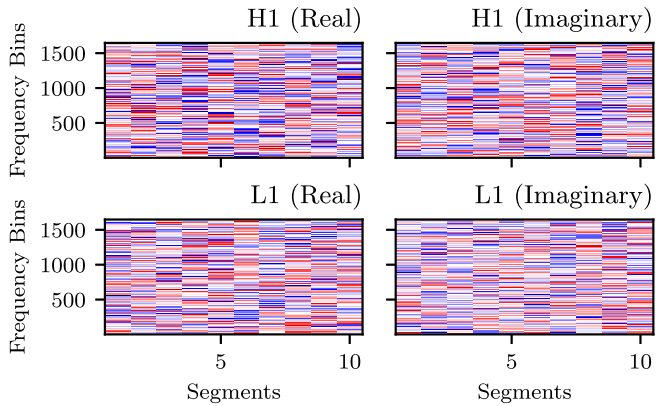

FIG. 4. Input spectrogram array for signal Sky-B@1000 Hz of Fig. 4(b) added to Gaussian white noise at a matched-filtering depth $\mathcal{D}^{90\%}_{\text{Sky}-\text{B}} \approx 81.8/\sqrt{\text{Hz}}$.

Motivated by these priors, we use the following design principles to construct our CW-DNN architecture:

(1) avoid operations that lose information about the signal (such as max-/mean-pooling),
(2) combine *all* signal bins within the shortest layer pathway,
(3) use an input spectrogram adapted to the diurnal repeating shifted signal pattern.

The last point is probably not strictly necessary, and is intended to simplify the problem for the network, by providing a "natural" factorization into a repeating shifted daily pattern that can be learned by the same convolutional kernels across all segments, producing a tracklike output pattern to be combined by subsequent layers.

### B. Preprocessing the input

We convert the one-dimensional time-series data $x(t)$ of Eq. (4) for each detector into a two-channel (real and imaginary part) spectrogram over ten one-day segments. Detectors are stacked along the channel dimension, and for two detectors (H1+L1) the input spectrograms therefore have a total of four channels. Hence, the input consists of a three-dimensional array, with axes corresponding to segments, frequency bins, and channels, as depicted in Figs. 3 and 4.

The input frequency band encompasses the entire signal bandwidth, aligned to start at the lowest signal frequency with a fixed total bandwidth corresponding to the widest signal in Table II, specifically ∼18.7 mHz for the Sky-B@1000 Hz case. With the segment FFT resolution of 1/day, plus a padding of 16 frequency bins on either side of the band, this results in a total input bandwidth of 1647 frequency bins.

### C. Network architecture

Through extensive experimentation based on the architecture design principles of Sec. IV A, we ultimately arrived at the simple network architecture summarized in Table III.

The first layer performs 1D-convolutions with 64 kernels of dimension $1 \times 313 \times 2$, sliding along the frequency-axis for each detector and segment. The kernel size of 313 frequency bins encompasses the widest signal bandwidth within the one-day segments, namely 3.4 mHz, for the skyB@1000 Hz signal.

The second layer performs 2D convolution of 64 kernels with dimension $2 \times 40 \times 64$, combining neighboring segments over 40 units along the frequency-axis. The width in frequency covers the widest output "track" width over the two-day span. This choice is motivated by the idea of combining the full signal information within the shortest possible network path, as discussed in Sec. IV A.

The output block consists of three layers: a flatten layer (reshaping the input to a one-dimensional array), a dense layer with 32 units and a final output layer consisting of a single unit.

Every layer except the flatten and output layers use ReLU activation. The output layer uses a sigmoid activation in the final output for the probability $\hat{y} \in [0, 1]$ of the data containing a signal.

TABLE III. DNN architecture for targeted ten-day CW search: the output shape (T, F, C) of each layer corresponds to the number of bins in the (time, frequency, channels) axes, respectively. The kernel sizes of the convolutional layers are using the same convention.

| Layer | Output shape (T, F, C) |
| --- | --- |
| Input | (10, 1647, 4) |
| Conv1D (Stride—16) | (10, 103, 64) |
| Conv2D (Stride—(1, 4)) | (10, 26, 64) |
| Flatten | (16640) |
| Dense | (32) |
| Output | (1) |

---

[2]This assumes *isolated* sources and needs to be revisited for sources in binary systems.

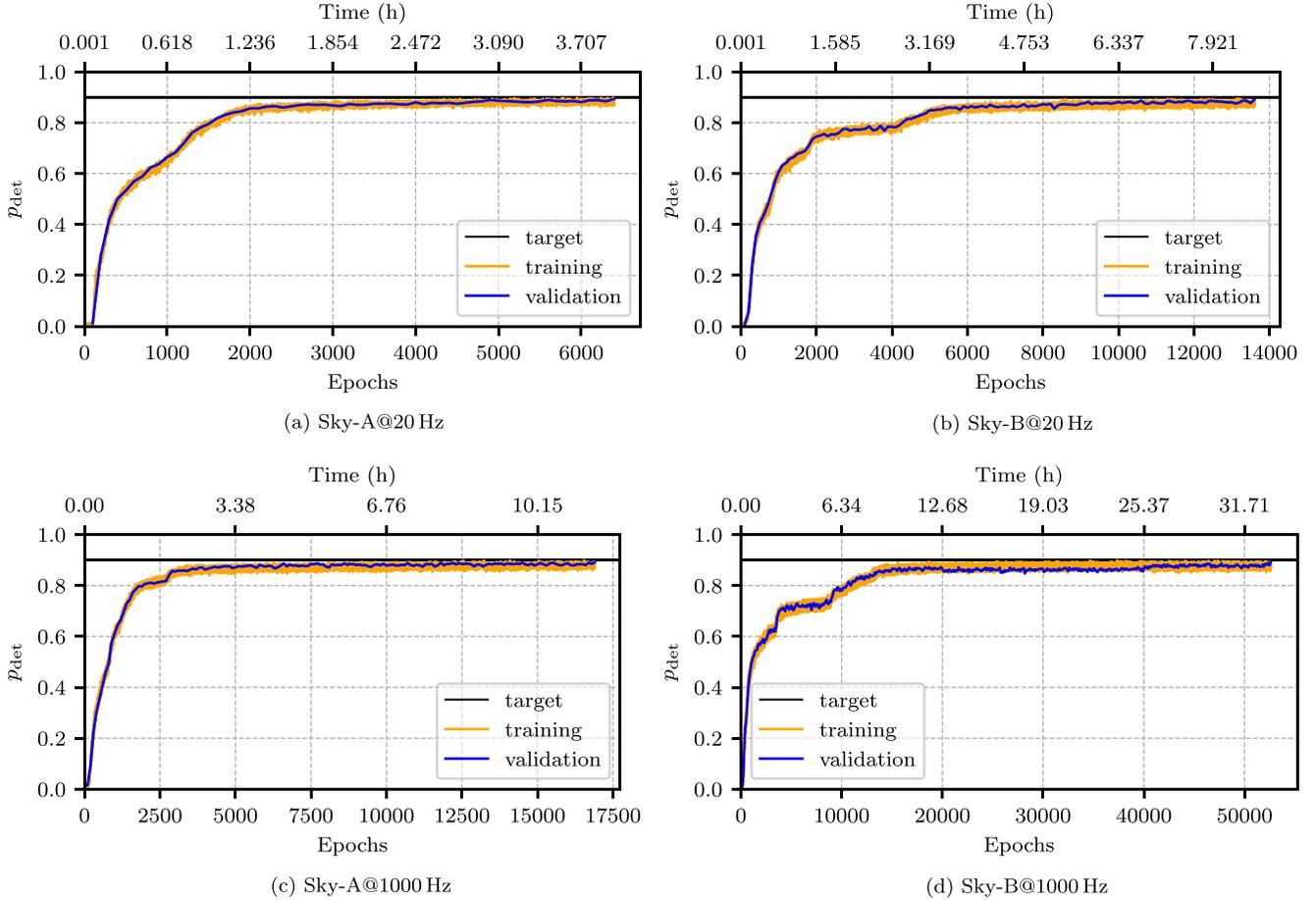

FIG. 5. Training and validation detection probability $p_{\rm det}$ (at fixed $p_{\rm fa} = 1\%$) versus number of epochs and training time for the targeted-search cases (a) Sky-A@20 Hz, (b) Sky-B@20 Hz, (c) Sky-A@1000 Hz, and (d) Sky-B@1000 Hz. See Table I for the complete parameter definitions and Table II for the corresponding signal bandwidths.

The sigmoid output $\hat{y}$ is well-suited for training a classification network, but as previously observed [2,21], this tends to run into numerical over and underflow issues when using it as a detection *statistic*, e.g., when measuring the receiver-operator-characteristic (ROC) of detection probability $p_{\rm det}$ versus false-alarm $p_{\rm fa}$. This can be avoided simply by dropping the final sigmoid activation when using the trained network's output as a detection statistic.

The total number of trainable parameters of the network in Table III is ∼900 k, and the network requires ∼5 MB of GPU memory per sample. The training was performed on NVIDIA A100-SXM4 GPUs with 40 GB of memory. The DNN was implemented in TENSORFLOW 2.0 ([22]) with the Keras API ([23]). We used the Weights and Biases platform [24] to track our experiments and to log training metrics.

Note that despite this being a rather shallow five-layer network, it still contains about half the trainable parameters of the nearly 100-layer deep Inception-Resnet-v2 of [2], which is due to the fact that our network uses substantially larger kernels.

### D. Training and validation

For each of the targeted-search cases in Table I we train the same network architecture on samples containing either pure Gaussian noise or an additional signal. The training data is constructed from a fixed set of 8192 precomputed (for performance reasons) signals with randomly-chosen amplitude parameters according to the physical uniform priors $\cos\iota \in [-1, 1]$, $\psi \in [-\pi/4, \pi/4]$ and $\phi_0 \in [0, 2\pi]$. Each signal is added to a dynamically-generated noise realization, at a fixed matched-filtering sensitivity depth of $\mathcal{D}^{90\%}$ as described in Sec. III.

In every epoch, the network is trained on all 8192 signals added to Gaussian noise and an equal number of pure Gaussian-noise samples, where the noise is dynamically generated in every sample. The training therefore never sees the exact same samples twice and therefore there can no overfitting or memorization in the strict sense, although the finite selection of 8192 signals from the continuous distribution can still result in some bias or small overfitting.

We use a binary cross-entropy loss function, which is common practice for classification tasks, namely

$$\mathcal{L}(y, \hat{y}) = \frac{1}{N} \sum_{i=1}^{N} -y^i \log \hat{y}^i - (1 - y^i) \log(1 - \hat{y}^i), \quad (7)$$

where $\hat{y}^i \in [0, 1]$ is the DNN sigmoid output for the $i^{\text{th}}$ sample, $y^i$ is the corresponding label (0 for noise and 1 for a signal), and $N$ is the total number of samples in a batch. We use the Adam optimizer with a batch size of 128 samples for training.

At every epoch, we measure the DNN detection probability $p_{\text{det}}$ at a constant $p_{\text{fa}} = 1\%$ false-alarm level on the training dataset. Every 100 epochs, we perform a validation step, where loss and detection probability are evaluated on an independent dataset drawn from the same distribution, constructed again from a (different) set of 8192 independent precomputed signals.

The learning progress of detection probability versus training epoch and time are shown in Fig. 5 for four representative cases. The network training continues until the validation detection probability exceeds $p_{\text{det}} \geq 89\%$. For each of the targeted-search cases, we start training from ten different random DNN weight initializations, and we use the best-performing network from each case for final testing.

These results confirm an empirical observation mentioned in Sec. III, namely the time required for the DNN to achieve matched-filtering performance increases with signal bandwidth, suggesting that is more "difficult" for the network to learn wider signals.

## V. RESULTS AND DISCUSSION

### A. Verifying performance on a test dataset

The close agreement observed in Fig. 5 between the DNN performance on the training and validation datasets indicates that there is no overfitting to the finite set of 8192 training signals. However, there is still potential for overfitting to the validation set during the optimization of the network hyperparameters (i.e., learning rate, layers, kernel sizes, strides, etc).

Therefore we evaluate the full-trained DNN on a completely independent *test* dataset. We generate new samples of Gaussian white noise and add signals at fixed matched-filtering depth $\mathcal{D}^{90\%}$ of Eq. (5) with randomly-drawn amplitude parameters $\cos \iota, \psi, \phi_0$, resulting in the final test detection probabilities $p_{\text{det}}$ at fixed false-alarm of $p_{\text{fa}} = 1\%$ shown in Table IV. These results are consistent with the validation detection probability of $p_{\text{det}} \geq 89\%$ that was used as a stopping criterion for the training. There is a slight downward bias of the test results, i.e., $\bar{p}_{\text{det}} \sim 88.3\%$, which makes sense given that training was stopped as soon as $p_{\text{det}}$ exceeded 89% on the validation dataset that is subject to both finite-sampling uncertainties and biases.

TABLE IV. Detection probabilities $p_{\text{det}}$ (at fixed $p_{\text{fa}} = 1\%$) achieved by the trained DNNs, evaluated on an independent *test* dataset for each of the five frequencies and the two sky positions Sky-A and Sky-B, see Table I for the complete parameter definitions.

| Frequency | Sky-A | Sky-B |
|---|---|---|
| 20 | $89.0^{+0.8}_{-1.2}$ | $88.5^{+0.8}_{-1.0}$ |
| 100 | $87.8^{+0.8}_{-1.1}$ | $87.4^{+1.0}_{-1.0}$ |
| 200 | $89.0^{+0.8}_{-1.0}$ | $89.0^{+0.9}_{-1.0}$ |
| 500 | $88.4^{+0.7}_{-1.0}$ | $88.8^{+1.0}_{-0.9}$ |
| 1000 | $87.6^{+0.8}_{-1.1}$ | $87.6^{+1.0}_{-1.2}$ |

### B. Detection efficiency versus signal depth

All results presented up to this point refer to signals at a fixed matched-filtering depth $\mathcal{D}^{90\%}$ of Eq. (5), i.e., signals with a fixed amplitude $h_0$. A valid question therefore arises if the network correctly generalizes to other signal amplitudes, as it could in principle have memorized or specialized to this particular sensitivity depth.

We measure $p_{\text{det}}$ of the trained DNN at varying signal depths $\mathcal{D}$, commonly referred to as the *efficiency curve*, shown in Fig. 6 for the Sky-B@1000 Hz case (results for the other test cases look similar). We can see that the DNN statistic behaves very similarly to matched filtering for both weaker and stronger signals compared to the $\mathcal{D}^{90\%}$ depth it was trained at. This confirms similar results found previously in [1,2], namely fixing the training depth to $\mathcal{D}^{90\%}$ does not seem to result in any over-specialization of the network.

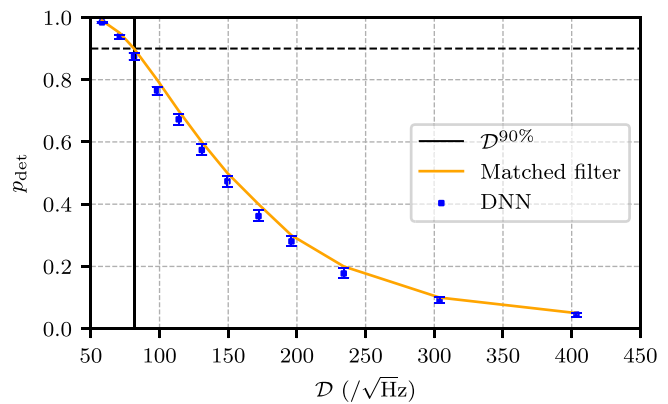

FIG. 6. Detection probability $p_{\text{det}}$ (at fixed $p_{\text{fa}} = 1\%$) versus signal depth $\mathcal{D}$ for the trained DNN (circles with 90% error bars) compared to matched filtering (solid line), for the benchmark case Sky-B@1000 Hz.

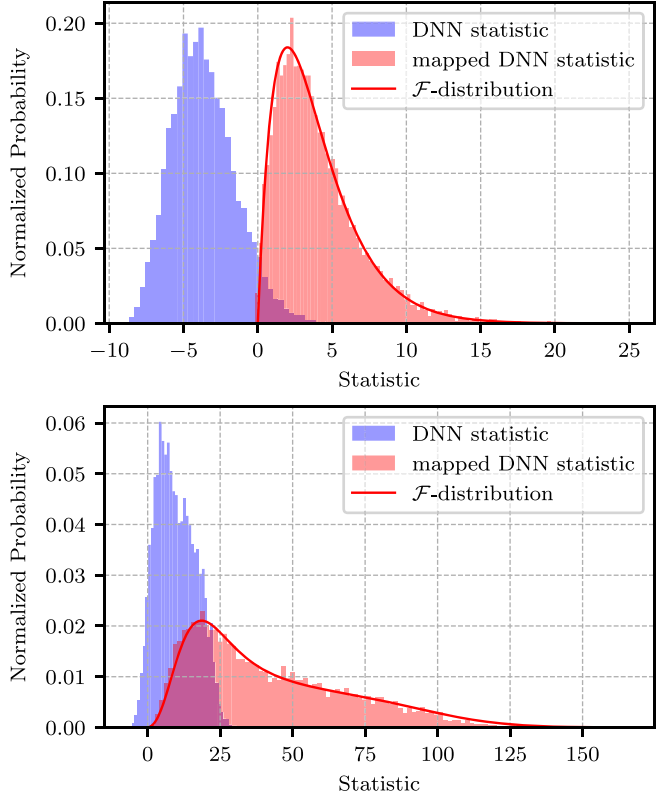

FIG. 7. Comparison between the distribution of the $\mathcal{F}$-statistic (solid line), the DNN output statistic (blue histogram) and the mapped DNN output statistic (red histogram) for pure noise samples (top) and $\mathcal{D}^{90\%}$ signals in noise (bottom), for the Sky-B@1000 Hz test case.

### C. Approximate mapping to the $\mathcal{F}$-statistic distribution

The results in the previous sections suggest that the receiver-operator-characteristic (ROC) $p_{\text{det}}(p_{\text{fa}})$ of the DNN statistic agrees well with that of the matched-filtering $\mathcal{F}$-statistic [14]. We therefore expect the DNN statistic and the $\mathcal{F}$-statistic to be (approximately) related by a monotonic function.

However, for simplicity here we only consider the corresponding statistic *distributions* (which fully determine their respective ROC performance), which would be related by the same monotonic function. We test this prediction by comparing the DNN statistic distributions to the known $\chi^2$-distribution of the $\mathcal{F}$-statistic[3] in both the pure noise as well as the signal + noise cases. For illustration purposes we focus on the "hardest" Sky-B@1000 Hz test case.

---

[3]The actual Neyman-Pearson-Searle optimal statistic [25] is not the $\mathcal{F}$-statistic but the Bayes factor, but for our present purposes the difference in ROC performance will be negligible [18].

We obtain a distribution of DNN output statistics on pure noise samples and fit a quadratic mapping to the known $\mathcal{F}$-statistic noise distribution, namely a central $\chi^2$-distribution with four degrees of freedom. The best-fit quadratic is obtained as $0.1x^2 + 1.9x + 9.2$ which is a monotonic function in the range of the DNN statistic. The resulting mapped noise distributions are shown in the top panel of Fig. 7.

We then apply the same mapping to the DNN statistic outputs obtained in the signal case with signals injected at depth $\mathcal{D}^{90\%}$, and we compare the resulting distribution to the corresponding fixed-depth $\mathcal{F}$-statistic distribution (cf. [16]), shown in the bottom panel in Fig. 7. We see reasonably good agreement between the DNN- and $\mathcal{F}$-statistic distributions, respectively, as expected. This shows (somewhat complementary to Fig. 6) that the network has properly generalized by learning to compute a statistic with similar characteristics to (near-)optimal "matched filtering" statistics such as the $\mathcal{F}$-statistic.

## VI. CONCLUSIONS

State-of-the-art convolutional image-classification networks have proven ineffective [1,2] for CW searches on longer durations of ∼11.6 days. We hypothesize that this failure is due to an inherent mismatch between the CW signal morphology and the priors implicit in (convolutional) image-classification network architectures.

We propose new DNN architecture design principles for CWs, which lead us to a novel convolutional DNN architecture that can effectively achieve matched-filtering sensitivity for targeted CW searches over ten days.

Future work needs to extend this study to longer durations (up to 1–2 years) and CW sources in binaries that would be subject to even larger Doppler spreads. The resulting network input sizes will become substantially larger as a result, potentially creating memory and performance bottlenecks. Furthermore, returning to wide-parameter-space searches will require scaling up the network *capacity* in order to be able to learn large numbers of different signal shapes.

More work will therefore be required to further improve the network architecture, for example, using transformers [26,27] for the 2D "track" processing (see Sec. IV C) might be an interesting direction with the potential of minimizing the network pathway combining the full signal power.

## ACKNOWLEDGMENTS

This work has utilized the ATLAS computing cluster at the MPI for Gravitational Physics, Hannover, and the HPC system Raven at the Max Planck Computing and Data Facility.